\documentclass[sigconf]{acmart}

\newcommand{\modified}[1]{\textcolor{black}{#1}}

\newcommand{\system}{EventAnchor}

\AtBeginDocument{%
  \providecommand\BibTeX{{%
    \normalfont B\kern-0.5em{\scshape i\kern-0.25em b}\kern-0.8em\TeX}}}

\setcopyright{acmcopyright}
\copyrightyear{2021}
\acmYear{2021}
\setcopyright{acmcopyright}\acmConference[CHI '21]{CHI Conference on Human Factors in Computing Systems}{May 8--13, 2021}{Yokohama, Japan}
\acmBooktitle{CHI Conference on Human Factors in Computing Systems (CHI '21), May 8--13, 2021, Yokohama, Japan}
\acmPrice{15.00}
\acmDOI{10.1145/3411764.3445431}
\acmISBN{978-1-4503-8096-6/21/05}

\acmSubmissionID{8268}

\begin{document}

\title{EventAnchor: Reducing Human Interactions in Event Annotation of Racket Sports Videos}

\author{Dazhen Deng}
\email{dengdazhen@zju.edu.cn}
\orcid{0000-0002-9057-8353}
\affiliation{%
  \institution{The State Key Lab of CAD\&CG, Zhejiang University}
  \city{Hangzhou}
  \state{Zhejiang}
  \country{China}
  \postcode{310000}
}

\author{Jiang Wu}
\email{wujiang5521@zju.edu.cn}
\affiliation{%
  \institution{The State Key Lab of CAD\&CG, Zhejiang University}
  \city{Hangzhou}
  \state{Zhejiang}
  \country{China}
  \postcode{310000}
}

\author{Jiachen Wang}
\email{wangjiachen@zju.edu.cn}
\affiliation{%
  \institution{The State Key Lab of CAD\&CG, Zhejiang University}
  \city{Hangzhou}
  \state{Zhejiang}
  \country{China}
  \postcode{310000}
}

\author{Yihong Wu}
\email{wuyihong@zju.edu.cn}
\affiliation{%
  \institution{The State Key Lab of CAD\&CG, Zhejiang University}
  \city{Hangzhou}
  \state{Zhejiang}
  \country{China}
  \postcode{310000}
}

\author{Xiao Xie}
\email{xxie@zju.edu.cn}
\affiliation{%
  \institution{The State Key Lab of CAD\&CG, Zhejiang University}
  \city{Hangzhou}
  \state{Zhejiang}
  \country{China}
  \postcode{310000}
}

\author{Zheng Zhou}
\email{zheng.zhou@zju.edu.cn}
\affiliation{%
  \institution{Department of Sport Science, Zhejiang University}
  \city{Hangzhou}
  \state{Zhejiang}
  \country{China}
  \postcode{310000}
}

\author{Hui Zhang}
\email{zhang_hui@zju.edu.cn}
\affiliation{%
  \institution{Department of Sport Science, Zhejiang University}
  \city{Hangzhou}
  \state{Zhejiang}
  \country{China}
  \postcode{310000}
}

\author{Xiaolong (Luke) Zhang}
\email{lzhang@ist.psu.edu}
\affiliation{
    \institution{College of Information Sciences and Technology, Pennsylvania State University}
    \country{United States of America}
}

\author{Yingcai Wu}
\email{ycwu@zju.edu.cn}
\authornote{Yingcai Wu is the corresponding author.}
\affiliation{%
  \institution{The State Key Lab of CAD\&CG, Zhejiang University}
  \city{Hangzhou}
  \state{Zhejiang}
  \country{China}
  \postcode{310000}
}
\renewcommand{\shortauthors}{Deng, et al.}

\begin{abstract}
The popularity of racket sports (e.g., tennis and table tennis) leads to high demands for data analysis, such as notational analysis, on player performance. While sports videos offer many benefits for such analysis, retrieving accurate information from sports videos could be challenging. In this paper, we propose EventAnchor, a data analysis framework to facilitate interactive annotation of racket sports video with the support of computer vision algorithms. Our approach uses machine learning models in computer vision to help users acquire essential events from videos (e.g., serve, the ball bouncing on the court) and offers users a set of interactive tools for data annotation. An evaluation study on a table tennis annotation system built on this framework shows significant improvement of user performances in simple annotation tasks on objects of interest and complex annotation tasks requiring domain knowledge.

\end{abstract}




\maketitle

\section{Introduction}
Racket sports are popular over the world. For example, tennis, often regarded as the top 1 racket sport, has more than 87 million players in 2019~\cite{itfreport2019}, and ATP (the Association of Tennis Professionals) events have attracted 1 billion cumulative viewers~\cite{sportseconomics}. Such popularity leads to high demands for data analysis on player performance by both amateurs and professional analysts~\cite{Lees2003, wu2020visual}.
One widely used analytical method for racket sports is notational analysis \cite{NotationalAnalysis,Lees2003}, which focuses on the movements of players in a match. Video recordings of matches are often used for such analysis because of the availability of rich source information, such as the position and action of players, action time, and action result. Manually retrieving massive source information from long match videos could be very challenging for users, so computer vision algorithms have been applied to data extraction from sports videos.

Existing data acquisition systems based on computer vision have several limitations. First, many systems cannot accurately track data from low-quality videos, such as broadcasting videos~\cite{thomas2017computer}. For example, the low frame rate in broadcasting videos cannot exhibit the fast motion of players and ball/shuttlecock well. In elite table tennis matches, where the average duration between two strokes can be as fast as just half a second~\cite{Lees2003, loh2015competition}, the image of the ball in a single frame may be a semi-transparent tail to show a series of ball positions. Most computer vision models cannot accurately recognize the ball position from the tail. Similarly, other characteristics in racket sports videos, including but not limited to the frequent shot transformation, inconsistent scene appearances, and severe occlusion of the ball/shuttlecock by players, also pose challenges for robust object recognition. Second, existing systems largely focus on low-level object recognition, such as human action~\cite{zhu2006action}, and are weak to identify and retrieve high-level event information, such as the outcomes of the actions~\cite{decroos2019actions}. This limitation is still an open problem in computer vision~\cite{kong2018human, thomas2017computer}, because automatically extracting contextual information in sports requires the integration of domain knowledge into algorithms.

Interactive data acquisition systems have been developed to improve the accuracy and quality of data extraction from videos. Such systems allow user involvement in the data processing, such as manually validating the tracking result of the ball or labeling the outcome of a serve. One of the challenges such systems face is the scalability. When having a large number of annotations to process, existing systems often rely on crowdsourcing~\cite{vondrick2013efficiently, papadopoulos2017extreme, krishna2016embracing}. Another challenge is annotation efficiency for individual users. Some research attempted to reduce human interaction in data annotation from sports videos, such as baseball videos\cite{ono2019historytracker}, but their methods cannot be applied to racket sports, which with faster and more dynamics rhythms, require different approaches for data annotation.

In this paper, we propose \system{}, an analytical framework to support data annotation for racket sports videos. Our framework integrates computer vision models for scene detection and object tracking, and uses the model outputs to create a series of anchor points, which are potential events of interest. Interacting with these anchors, users can quickly find desired information, analyze relevant events, and eventually create annotations on simple events or complex player actions. Based on the framework, we implement an annotation system for table tennis.
The results of our evaluation study on the system show significant improvement of user performances in data annotation with our method. 

\modified{
The major contribution of this paper lies in the novel framework, \system{}, that we propose for multiple-level video data annotations based on our empirical work in understanding the requirements of data annotation by expert analysts. This framework integrates rich information and supports efficient video content exploration.}

\section{Related Work}
Our research focuses on interactive video annotation enhanced with machine learning techniques in computer vision. Thus, in this section, we review the methods for video annotation, particularly those relying on machine-learning or crowdsourcing to scale up annotation. We also discuss research on interaction design to support video annotation.

\subsection{Model-assisted Video Annotation}
The advance of machine learning has provided new opportunities to reduce the cognitive and interaction burdens of users in video annotation\cite{liao2016visualization, heilbron2018annotate, vondrick2011video, song2019popup}. Models have been incorporated into video annotation systems for various purposes, such as predicting annotations based on user interaction activities ~\cite{vondrick2011video,heilbron2018annotate,liao2016visualization}, and propagating the annotation of keyframes to other frames~\cite{wang2007automatic,lavrenko2004statistical}. Many different models have been considered. For example, the models to predict annotation include those based on continuous relevance~\cite{lavrenko2004statistical}, particle filtering\cite{xie2020passvizor, wu2018forvizor}, and bayesian inference~\cite{wang2007automatic}. 
One common approach in model-based video annotation is to pre-process data with models pre-trained with other datasets. This practice can improve the efficiency of data annotation by removing non-interesting data. For example, when constructing the NCAA Basketball Dataset, Ramanathan et al.~\cite{ramanathan2016detecting} used a pre-trained classifier to filter video clips first, so that those non-profile shots can be eliminated before distributing the data and tasks to crowd workers. This approach can significantly reduce the amount of data for annotation, as well as the burdens of users in annotation. 

Motivated by these methods, this research uses computer vision models to extract essential entities and objects from racket sports videos, such as key frames, ball trajectories, and player positions. Despite the inevitable errors accompanied with such models, these entities and objects lay the foundations for further data processing (e.g., event recognition), and user interaction (e.g., searching and evaluating events of interest), therefore potentially improving annotation efficiency.

\subsection{Interaction Design to Support Video Annotation}
Researchers have also explored ways to help people annotate video data through interaction designs. One research direction is to explore new interactive approaches to facilitate important annotation tasks, such as an adaptive video playback tool to assist quick review of long video clips~\cite{higuchi2017egoscanning, joshi2015real, al2014fast}, a mobile application to support real-time, precise emotion annotation~\cite{zhang2020rcea}, an interaction pipeline for the annotation of objects and their relations~\cite{shang2019annotating}, and a novel method to acquire tracking data for sports videos~\cite{ono2019historytracker}. These designs, which largely target single users, can improve the efficiency and accuracy of video annotation from different perspectives. 
\modified{
Our proposed method is different from the aforementioned works from two perspectives. First, our method allows users to locate events of interest by integrating not only essential information at the object level (e.g., ball position), which existing designs~\cite{higuchi2017egoscanning, joshi2015real} largely focused on, but also more advanced information at the event (e.g., stroke type) and context (e.g., tactical style) levels, which we propose to enable more comprehensive and in-depth data analysis. Second, our method supports a more efficient and scalable exploration of events with computer vision algorithms and an improved timeline tool. Our algorithms can remove useless contents and keep the key events to better support fast and dynamic video review. Our fine-grained timeline, which visualizes the events at the frame level and is controlled by a calibration hotbox, allows users to quickly examine frames back and forth, even in very long videos.}

\begin{figure*}[tb]
    \centering
    \includegraphics[width=0.9\textwidth]{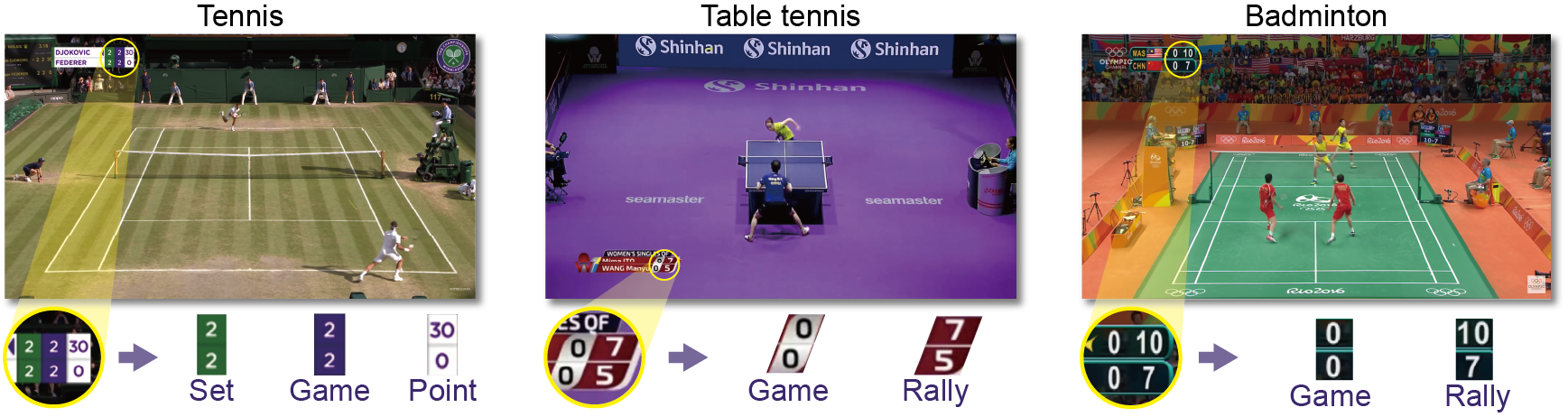}
    \caption{The snapshots of the broadcast videos and the scoreboards of tennis, table tennis, and badminton, respectively.}
    \label{fig:broadcast}
    \Description[Snapshots of broadcast videos of different racket sports]{The snapshots of the broadcast videos and the scoreboards of tennis, table tennis, and badminton, respectively.}
\end{figure*}

Another research direction focuses on designs to support crowd workers. Crowdsourcing has been considered as a way to scale up interactive annotation~\cite{luz2015survey, yuen2011survey}. While some work studied general design issues, such as user interface design guidelines for crowd-based video data annotation~\cite{kovashka2016crowdsourcing},  most research in this direction explored designs to combine annotations from the crowd to generate better results. For example, Kaspar et al.~\cite{kaspar2018crowd} developed an ensemble method to improve the quality of video segmentation, and Song et al.~\cite{song2019popup} proposed an intelligent, human-machine hybrid method to combine crowd annotations for 3D reconstruction. In addition to these works from a technical design perspective, some research also investigated non-technical issues in the design of crowdsourcing tools, such as the skills and motivation of crowd workers~\cite{vondrick2013efficiently}, and workflow for crowd workers~\cite{kim2014crowdsourcing}.
For sports videos, while most work used crowdsourcing to enhance data analysis~\cite{van2012automatic, perin2013real} or model training~\cite{ramanathan2016detecting}, Tang et al.~\cite{tang2012epicplay} developed a crowdsourcing method to construct annotation for video highlights based on social media data from sports fans. \modified{In this work, we focus on improving the efficiency of single workers.}

\subsection{Video Annotation Software}
Various tools~\cite{doermann2000tools, yuen2009labelme, vondrick2013efficiently, biresaw2016vitbat, bianco2015interactive} have been developed for video data annotation. Early work largely focused on object recognition and annotation. For example, ViPER~\cite{doermann2000tools} can annotate the bounding boxes of objects and texts frame by frame, and LabelMe~\cite{yuen2009labelme} supports the annotation of the same object across different frames. As the demands for video annotation dramatically increased, efforts were made to reduce the burdens in the annotation. VATIC~\cite{vondrick2013efficiently}, for example, was designed to leverage crowdsourcing for video annotation; iVAT~\cite{bianco2015interactive} combined automatic label generation with user manipulation to improve annotation efficiency; ViTBAT~\cite{biresaw2016vitbat} supported the annotation of individual and group behaviors across different frames.

These projects laid the foundations for the design of video annotation systems. Some methods, such as the bounding box in ViPER and frame interpolation in LabelMe, have become common practices supported by many annotation tools. Basic functions like geometry drawing (e.g., lines, rectangles, polygons) and video operation (e.g., pause, speed control, skip back or forward) have been widely adopted. However, these tools only support basic annotation tasks, such as labeling objects from general videos, with limited support for annotation tasks involving multiple fast moving objects across space and temporal dimensions, as what racket sports videos usually have.
\section{Racket Sports and related analytical Problems} 
\label{background}
In this section, we first explain the major rules of racket sports and some characteristics of broadcasting videos that may affect data acquisition. Although a match in racket sports can be single or double competition, we use single matches as examples. Also, we focus on those typical racket sports with a net to separate players, such as tennis, table tennis, and badminton. Those sports in which players are not separated by a net and can have direct body contacts, such as racquetball and squash, are not considered because of the different video scene structures. We will also introduce our two studies to learn about the tasks and data in video annotation. The first study is an interview study with three domain experts. The second study is a survey investigation to collect information on the interests of sports fans.

\begin{table*}
  \caption{The common analytical tasks based on interview data.}
  \label{tab:question}
  \begin{tabular}{|ccc|l|l|}
    \toprule
    T & TT & B & Tasks & Level\\
    \hline
    \checkmark & \checkmark & \checkmark &T1. Who served?  (server) & Object\\
    \checkmark & \checkmark & \checkmark &T2. What was the type of serve? (serve type)& Context\\
    \checkmark & \checkmark & \checkmark &T3. What was the effect of serve? (serve effect)& Context\\
    \checkmark & \checkmark & \checkmark &T4. Where did the ball fall on the court? (ball position)& Event\\
    \checkmark  &  & \checkmark &T5. What was the speed of the ball (shuttlecock)? (ball speed)& Object\\
    \checkmark & \checkmark &   &T6. What was the spin type of the ball? (ball spin)& Context\\
    \checkmark & \checkmark & \checkmark &T7. How was the ball received? (receive type)& Context\\
    \checkmark & \checkmark & \checkmark &T8. What was the effect of receiving? (receive effect)& Context\\
    \checkmark & \checkmark & \checkmark &T9. Where was the server/receiver? (player position)& Event\\
    \checkmark & \checkmark & \checkmark &T10. How did the server/receiver move before/after hitting the ball? (player movement)& Event\\
    \checkmark &\checkmark &\checkmark &T11. Who won this rally? (rally winner)& Object\\
    \checkmark &\checkmark &\checkmark &T12. What was the tactic of the player in this rally? (rally tactic)& Context\\
  \bottomrule
  \multicolumn{4}{l}{Note: T---Tennis, TT---Table Tennis,  B---Badminton.}
\end{tabular}
\end{table*}

\subsection{Racket Sports and Match Broadcasting Videos}
\label{match}
The match structures of racket sports are similar. A \textbf{match} is a competition between two players. A match is usually played in the best of N (e.g., 3, 5, 7) \textbf{games}, and each game is played in the best of N \textbf{rallies} (or points). The only exception is tennis, where there is another layer called set above the game. Tennis is played in the best of N sets (Fig.~\ref{fig:broadcast}). When playing a rally, two players hit the ball (or shuttlecock) in turns until one fails to send the ball to the court on the other side and loses one point~\cite{Lees2003}. Each hit is called \textbf{a stroke}, and the first stroke in a rally is called the serve. 

Broadcasting videos of racket sports include different types of contents. The central piece is the rallies, which are shown without interruption and often with a fixed camera angle to ensure the coverage of the whole court, as shown in Fig.~\ref{fig:broadcast}. Before a rally, videos usually capture how players prepare for the rally (e.g., resting, chatting with coaches). After a rally, audience reactions often appear, and a rally replay in slow motion may also be provided. What are essential to data annotation are those rally segments, and the values of other contents are minimal. The duration of a rally varies from sport to sport, ranging from seconds to minutes~\cite{Lees2003}, but a match can last hours, as often seen in tennis. 

\subsection{Studies on Interests of People in Racket Sports}
We designed two studies to learn about how a match is analyzed and what data is used in the analysis. Considering the diverse interests of people and possible vast design space, we first conducted an interview study with domain experts to identify the essential tasks in analysis, data required by analysis, and common challenges in data acquisition. Based on the information collected from this study, we designed a survey to investigate what ordinary sports fans may be interested in, and what kinds of problems they may have had if they have been involved in data annotation. 

\subsubsection{Expert Interview}
\label{sec:interview}
Our interview study is a semi-structured investigation involving three domain experts: \textbf{E1, E2}, and \textbf{E3}. \textbf{E1}, a professor of sports science, is interested in table tennis analysis. 
\textbf{E2} is a badminton analyst and also a professor at a top sports university.
Both \textbf{E1} and \textbf{E2} have experience in data analysis for more than twenty years.
\textbf{E3} is a Ph.D. candidate of sports science, and as a former professional tennis player, has conducted research on tennis data analysis for more than three years. Our interviews with \textbf{E1} and \textbf{E3} were in a face-to-face manner, and the meetings with \textbf{E2} was through a real-time, video conference call.

Three interviews followed the same structure. Each interview had two sessions. The questions in the first session were the same for all three experts, and focused on the understanding of their analytical tasks and relevant data in their own domain. The conversations in the second session were based on the information gained from the first session, and aimed at deepening the understanding of the challenges in analysis and current approaches to address them. Each interview lasted about 90 minutes: roughly 60 minutes for the first session, and 30 minutes for the second.

In the first session, we learned about the commonality and uniqueness of analytical tasks in these sports. The interests of the three experts were almost the same at a high-level. They were all interested in analyzing the movement of the ball (shuttlecock), the movement of players, their tactics in a rally (e.g., the type and effect of a serve), and the outcome of each rally. However, for certain tasks, their focuses differ. For example, in the analysis of ball movement, ball speed is a major factor to tennis and badminton, not to table tennis, and ball spin type is very crucial to table tennis or tennis, not at all to badminton. What distinguishes their analyses most are their strategies. In table tennis, where a rally is usually very short, the analysis often focuses on the scoring rates of players in different stages of a rally~\cite{wu1989methods,zhang2018match} and the tactics used by a player (e.g., the stroke position of a player, the landing location of the ball, and the stroke type~\cite{wang2019tac, wu2017ittvis}). 
In comparison, in badminton, the strategy centers on the three-dimensional trajectory of the shuttlecock\cite{ye2020shuttlespace}, because it can fundamentally affect the tactics in both offense and defense.
In tennis, which has a much larger court and a larger ball than table tennis and badminton, managing the physical energy by predicting the ball position and moving in advance is critical to tennis players. Therefore, the analysis often emphasizes player movement and its correlation with ball position~\cite{hughes200236}, in order to understand the spatio-temporal shot patterns~\cite{polk2014tennivis, polk2019courttime} and how players use various techniques~\cite{zhang2013evaluation} to mobilize their opponents to move.

\begin{figure*}[!tb]
    \centering
    \includegraphics[width=\textwidth]{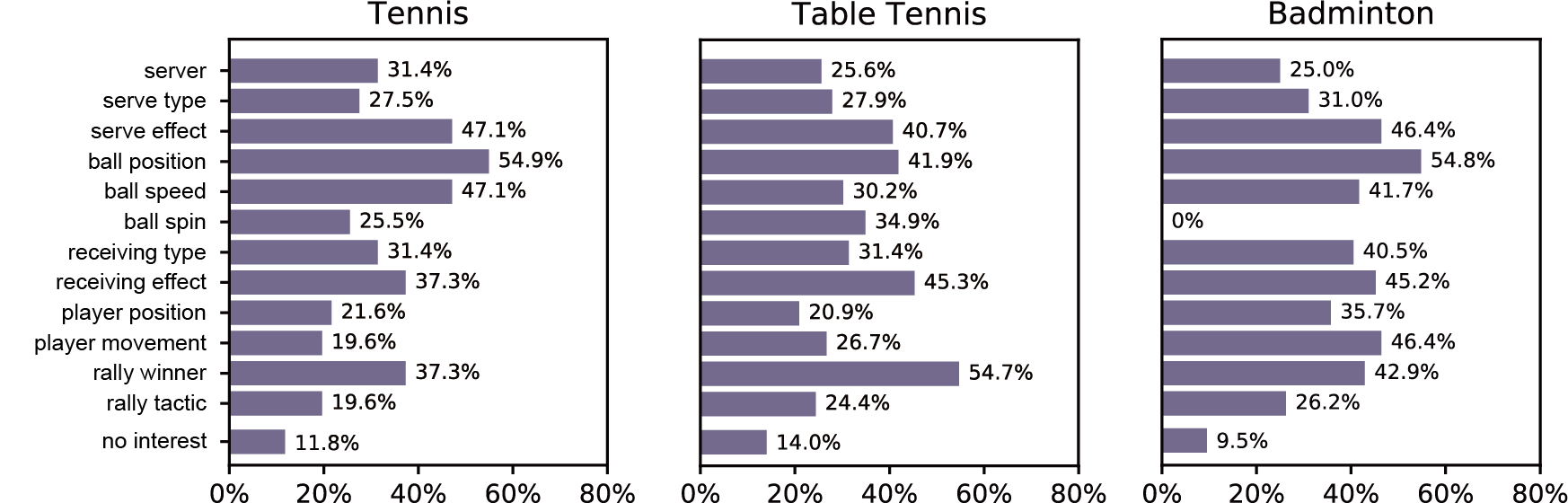}
    \caption{Tasks that sports fans are interested in when watching racket sports videos.}
    \label{fig:survey}
    \Description[Tasks that sports fans are interested in]{Tasks that sports fans are interested in when watching racket sports videos.}
\end{figure*}

In the second session, we gathered information about how these experts conducted their analysis. They all used certain software. However, their tools are usually very basic, largely limited to controlling video playback, capturing video images, and extracting video segments from a long video clip, and cannot support more advanced tasks, such as identifying important events, relating different events, and constructing annotations. For example, in table tennis analysis, \textbf{E1} usually needed to first specify the start time and end time of all rallies, and then drilled down into them to label \textit{ball position}, \textit{player position}, \textit{stroke type}, and \textit{spin type} of each stroke. However, searching the starts and ends of rallies through a long video is a tedious process, and manually clipping individual rallies out of the whole video is exhaustive. In addition, no tool is available for accurately specifying ball and player position. As a compromise, a common practice is to use a $3\times6$ grid on the virtual table to label the rough position of the ball, and four cells on each end of the table to indicate the area of player location.
Similar challenges also exist in tennis.
\modified{\textbf{E3} usually used a virtual court with a dense grid for the position of the ball and players.
In badminton, the three-dimensional trajectory of the shuttlecock is estimated by a physical motion model~\cite{Chen2009}.
To specify the three-dimensional start and end positions of the ball,  \textbf{E2} used a tool with a vertical view of the court for (x, y) coordinates and an end view for the z coordinate.
These tools were mostly developed in-house by their supporting staff, not commercially available.}

In addition, the experts encountered more challenges in those advanced tasks that require domain knowledge, such as identifying a stroke type in table tennis, which has to be inferred based on ball position and player position.
\modified{The video annotation tools help the experts to conduct opponent analysis and prepare players for their future matches. For example, knowing the tactics and strategies at different levels, players can take appropriate actions, such as avoiding those situations where the opponents have high winning rates. This group of users is the primary users of this research.}

Based on the data collected from the interview study, we summarized the primary tasks that are commonly seen in three sports, as shown in Table~\ref{tab:question}. Each task reflects a question that experts tend to ask in analysis. We use a simple term, which is inside the parentheses after each question, as a reference for each task. All tasks, except two, are interesting to all three. These two tasks are ball speed, which is not a concern in table tennis, and ball spin, which is not applicable to badminton. We still consider these two tasks in this research because they are very critical to other two sports.

\subsubsection{Survey Study}
We conducted a survey to learn what general sports fans may be interested in. The backgrounds and interests of sports fans could be very diverse. To keep our focus, we designed a questionnaire based on those tasks developed in the interview study.

The questionnaire includes demographic questions, task interest questions, and data annotation questions. Demographic questions sought some basic information from respondents related to their familiarity with and involvement in tennis, table tennis, and badminton, as well as their experiences in watching racket sports videos. Task interest questions were developed by drawing on the tasks in Table~\ref{tab:question}, and asked respondents which tasks they are interested in when watching match videos. In addition to these tasks, respondents could also choose none of these tasks and provide other tasks. Data annotation related questions asked whether respondents have been involved in data annotation for sports videos, and if so, what challenges they may have had.

We distributed the survey to two online communities in China. The total number of members in the two communities are more than 600. We got answers from 109 respondents. Among them, 51 (46.8\%) said they had watched tennis match videos, 86 (78.9\%) table tennis videos, and 85 (77.1\%) badminton videos.  

Most respondents indicated that they were interested in some of the tasks on the list (Figure~\ref{fig:survey}). Only a small portion of them showed no interest in any of them: 11.8\% in tennis responses, 14\% in table tennis, and 9.5\% in badminton. For tennis, the top three tasks are ball position (54.9\%), ball speed (47.1\%), and serve effect (47.1\%), and the least favourite tasks are two tied choices---player tactic (19.6\%) and player position (19.6\%). The top three tasks in table tennis are rally winner (54.7\%), receive effect (45.3\%), and ball position (41.9\%), and the bottom one is player position (21.0\%). The top three tasks in badminton are shuttlecock position (54.8\%), serve effect (46.4\%), and player position (46.4\%), and the least concern is who the server is in a rally (25\%). 

\begin{figure*}[tb]
    \centering
    \includegraphics[width=\textwidth]{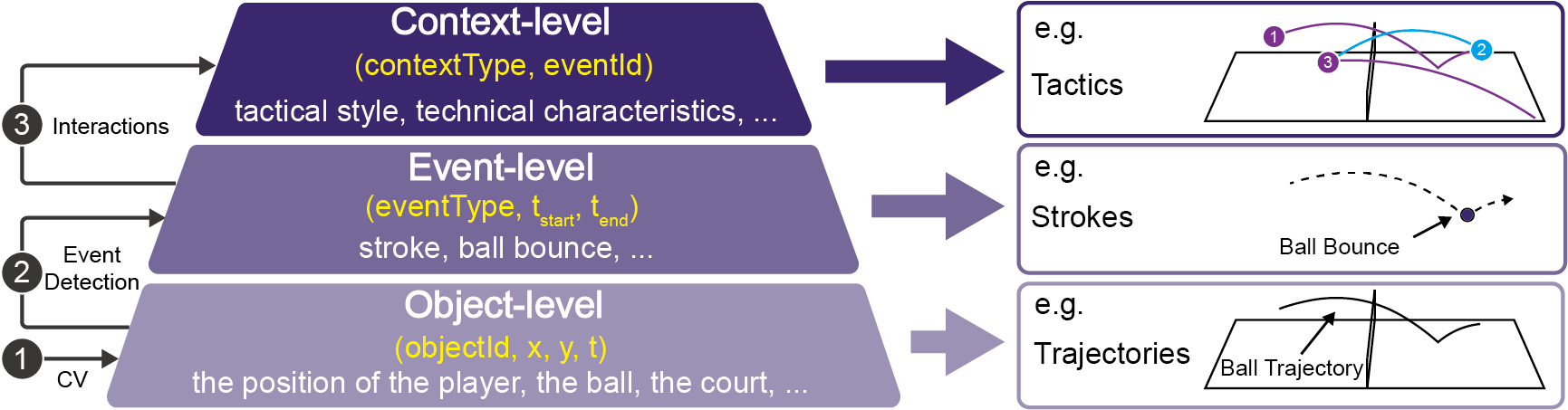}
    \caption{The \system{}  framework. Data at the object level includes objects recognized by computer vision algorithms. The event-level data is obtained through event detection algorithms based on the object-level data. The context-level data is results of user-machine collaboration, where users apply domain knowledge to select and integrate information from lower levels and and video.}
    \label{fig:framework}
    \Description[The EventAnchor  framework]{The EventAnchor framework consists of three levels of data, namely, object-level, event-level, and context-level.}
\end{figure*}

Only a few respondents had been involved in video data annotation. There are 10 people indicating experience in annotating table tennis videos, 2 in badminton, and 1 on tennis. One person had experience in all three. For table tennis annotation, two challenges stand out: accurately finding the times of important events (70\%) and locating a specific rally in a long video (60\%). Two challenges mentioned in annotating badminton videos are estimating shuttlecock location (100\%) and finding the times of important events (50\%). The only challenge given in tennis is locating a specific rally in a long video.

\modified{Novice users use video annotation tools differently from experts (Section \ref{sec:interview}). As our survey data shows that fans are more interested in events like who won a rally, where a ball landed, and how fast a fall was. We can speculate some application scenarios of our design by this group of users, such as using it to help the creation of highlight videos of a match or tutorial videos based on matches. The proposed method allows them to quickly identify and understand those key events in a match and choose their desirable video segments.}
\section{\system:  Supporting Multi-Level Video Annotation}
\label{methodology}

\system{} was developed based on literature on sports data analysis and what we learned from the interview and survey studies.
It has been argued~\cite{shih2017survey} that tasks in video analysis can involve information at different levels, ranging from raw objects (e.g., ball, court, player) at the bottom level to advanced inference or semantic analysis at the top level (e.g., player tactic). The primary tasks shown in Table~\ref{tab:question} actually include tasks at different levels. For example, some tasks like ball position, ball speed, and player position  are low-level tasks that concern object recognition, while tasks like serve  type, serve effect, and rally tactic are high-level semantic tasks that require domain knowledge to relate various aspects of spatial and temporal information about the ball and players. 

In-depth analysis of these tasks indicates that they are all related to a few key events: ball (shuttlecock)-racket contact and ball-court contact. For example, such tasks as server, serve type, receive type, and player position are all about situations before or after the event of ball-racket contact; and tasks like serve effect and receive effect are related to the ball-count contact event. Other advanced semantic tasks require the integration of information related to a series of such events. 

Based on this understanding, we develop a three-level framework, which has an event level in the middle to connect an object level below and a context level above (Fig.~\ref{fig:framework}). At the bottom is \textbf{the object level}. The data at this level is the foundation of the whole framework, and includes essential objects recognized by computer-vision algorithms from videos, such as the positions of the ball, the player, and the court. 
Data at this level can be represented as a tuple, $(objectId, x, y, t)$, where $objectId$ is the identity of a recognized object, $x$ and $y$ are the coordinates of the object in a video frame, and $t$ represents the timestamp of the frame where the object is.

The center level of the framework is \textbf{the event level}.  Data at this level concerns the interaction between essential objects from the object level, such as a stroke, which is the result of the ball contacting a racket, and the aggregation of them (e.g., a rally with multiple strokes). Data at this level comes from the information at the object level, such as the moving direction of the ball, or machine learning models that recognize events. The data can be represented as a tuple, $(eventType, t_{start}, t_{end})$, where $eventType$ represents the type of events, and $t_{start}$ and $t_{end}$ are the timestamps of the start and end of the event, respectively. 

Data at \textbf{the context level} summarizes information from the event level, and can include the technical attributes of strokes (e.g., stroke type, spin type) and the tactical style of a player.
Retrieving data at this level requires extensive annotation by domain experts, because of the required domain knowledge.
For example, to determine the type of a stroke in table tennis demands skills to recognize a sequence of micro-actions of the hand and wrist. Only analysts with extensive knowledge can make a right call. Similarly, obtaining the contextual information of some events, such as what player tactics a rally is based on, also requires domain knowledge. 
Data structure at this level can also be a tuple, $(contextType, eventId)$, where $contextType$ represents the type of context information and the $eventId$ the identity of the event. \modified{We provide a mapping from the analytical tasks to the data level in Table \ref{tab:question}.}

The event level plays an important role in this framework. Recognized events at this level are the anchors for analytical tasks. Knowing the locations of these events in a video, analysts can examine the images around them, find relevant video segments, and create corresponding annotations. For example, table tennis players prefer to launch an attack as early as possible in a rally, so analysts often want to examine those rallies in which a player launches an attack immediately after the serve and gains a point. To identify such rallies, users can rely on event information to select those short rallies as candidates and then apply domain knowledge to determine what rallies are of interest. Because of the essential role of the event level to link analytical tasks, we call the framework \system{}.

\section{Implementation of \system{} for Table Tennis}
\begin{figure*}[tb]
    \centering
    \includegraphics[width=\textwidth]{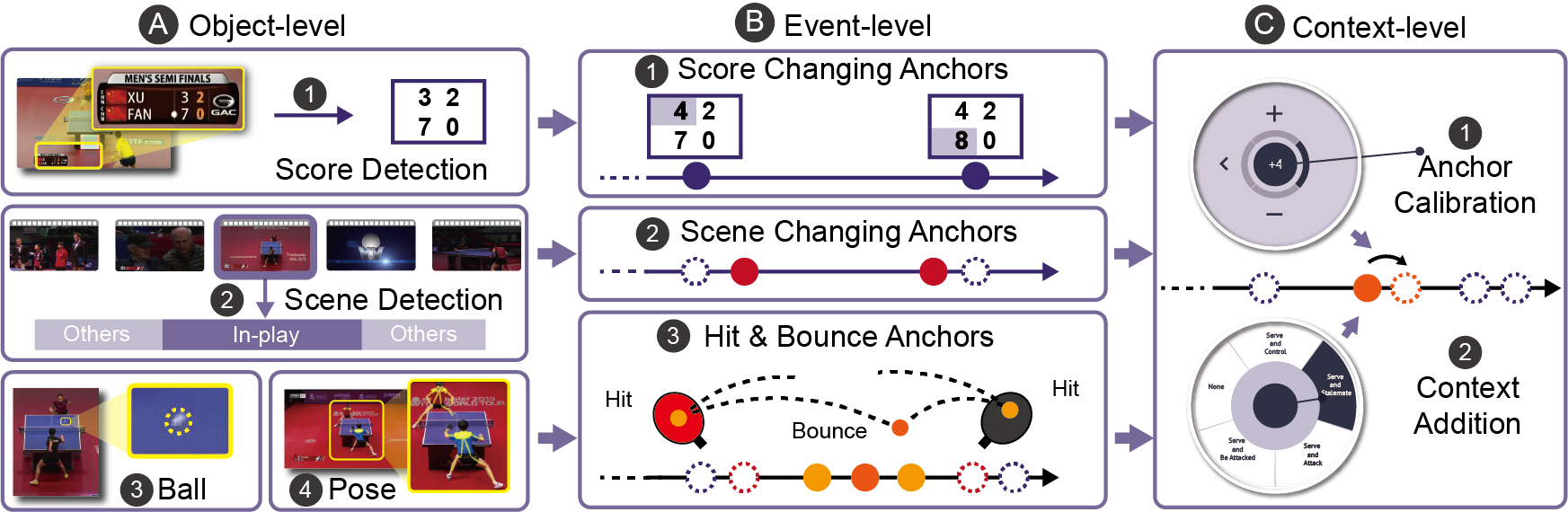}
    \caption{Pipeline of \system{} for Table Tennis. (A) exhibits the extraction of essential objects by computer vision models, such as score (A1), scene (A2), ball (A3), and player pose (A4); (B) describes the methods to obtain events by estimating the moment of score changing (B1) or scene changing (B2), or identifying the moments of ball hitting and ball bouncing (B3); and (C) shows two interactive tools for calibrating an event (C1) and annotating the event with contextual information (C2).}
    \label{fig:model}
    \Description[Pipeline of EventAnchor for Table Tennis]{The data processing pipeline of EventAnchor for table tennis is illustrated.}
\end{figure*}
Based on our framework, we implemented a system, \system{} for Table Tennis (ETT), to support annotation on table tennis videos. We chose table tennis because annotation on table tennis videos is often regarded as one of the most challenging tasks among racket sports. First, we used computer vision models, such as object detection~\cite{ren2015faster, Redmon2016YouOL}, object tracking~\cite{Wojke2017simple,sort16, huang2019tracknet}, and pose estimation~\cite{cao2018openpose, toshev2014deeppose} models, to identify the player, the ball, the court, and relevant trajectories (object-level) (Fig.~\ref{methodology}A).
Second, based on the motion of the ball and the player, as well as their relative position, we obtained events. The positions and timestamps of the events are used as anchors (event-level) (Fig.~\ref{methodology}B).
For example, a sudden change of the moving direction of the ball implies the event that a player hits the ball or the ball bounces on the table.
Anchors can help the user quickly locate individual events in videos.
Third, through visual interaction, the user can add contextual information (context-level), such as the technical characteristics of a stroke and the tactical style of a rally to each anchor, or calibrate the spatial and temporal information of an anchor (Fig.~\ref{methodology}C).

\begin{figure*}[!tb]
  \centering
  \includegraphics[width=\textwidth]{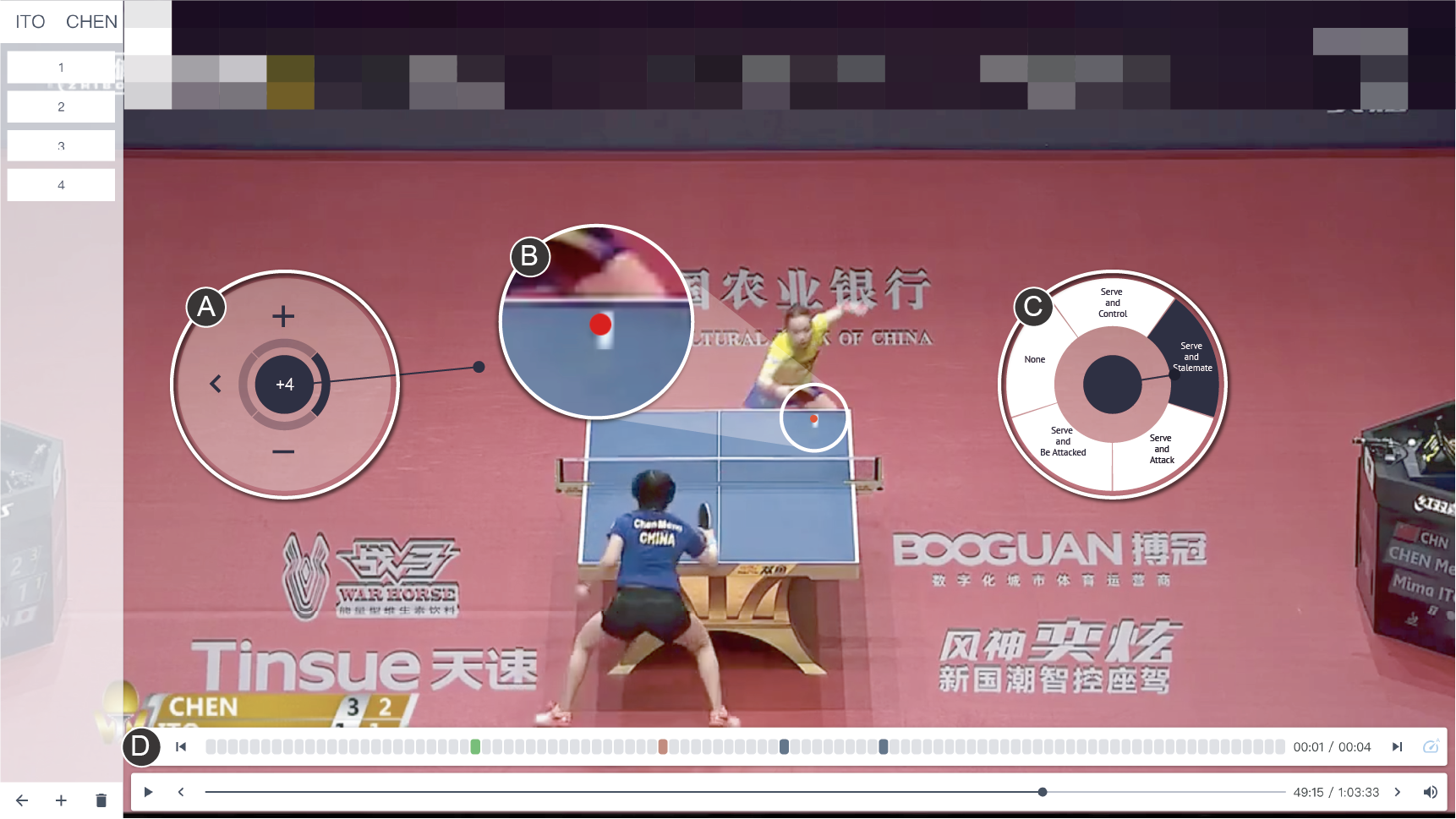}
  \caption{User interface of \system{} for Table Tennis. The interface includes three major components: anchors (B, D), a calibration box (A), and an annotation box (C). The figure shows a scenario where a user is correcting the timestamp of the second anchor with the calibration box.}
  \label{fig:interface}
  \Description[User interface of EventAnchor for Table Tennis]{The user interface of EventAnchor includes three major components: anchors, a calibration box, and an annotation box.}
\end{figure*}

\subsection{Acquiring Object-level Information}
To acquire object-level data from videos, we adopted a series of computer vision models~\cite{liu2018fots, he2015deep, huang2019tracknet, cao2017realtime}. Video processing had three steps: score detection, scene detection, and ball and pose recognition.
First, we used FOTS~\cite{liu2018fots}, an optical character recognition model, to process the scoreboard in video (Fig.~\ref{fig:model}A1).
We sampled 5,000 images from videos and annotated the location of the digits through crowdsourcing.
The retrieved images were separated into a training set (70\% of the whole data set) and a test set (30\% of the data set) for model training and test.
\modified{On the test set, the FOTS obtained a precision of 92.1\% and a recall of 95.4\%.}
Second, we classified the frames according to the scenes (Fig.~\ref{fig:model}A2).
Each frame was pre-processed with ResNet-50~\cite{he2015deep} that was pre-trained on the ImageNet~\cite{krizhevsky2012imagenet}, and an embedding vector with a length of 2,048 was obtained.
Given the embeddings, we conducted binary classification with support vector machine and obtained the frames of ``in-play.'' 
Third, to recognize the ball and player posture (Fig.~\ref{fig:model}A3,~\ref{fig:model}A4).
we used TrackNet~\cite{huang2019tracknet}, a ball tracking model for tennis and badminton, to extract ball trajectory.
\modified{By stacking three consecutive frames as the model input, the TrackNet can resolve the problems of noisy objects (e.g., white dots in the billboard or headband of the player being recognized as the ball), transparent tails, and invisible or severely blurred ball.}
To apply TrackNet in table tennis, we sampled over 60,000 frames from different videos to annotate ball positions.
After training, the TrackNet achieved an accuracy of 88.6\%.
For pose recognition, we used Openpose~\cite{cao2017realtime} trained on the COCO  dataset~\cite{lin2014microsoft}.

\subsection{Acquiring Event-level Information.} 
We used the object-level information to obtain anchors at the event level.
First, we segmented a video into a set of rallies by detecting the timestamps of score changes (Fig.~\ref{fig:model}B1). \modified{With the scores detected, we adopted the longest increasing sub-sequence algorithm to model score change and obtained the match structure. The accuracy of rally segmentation is 98.5\% in the test set.}
Second, based on the scene detection results, we derived the start and end frame of each rally (Fig.~\ref{fig:model}B2).
Third, combining the ball trajectory and player poses, we recognized the events such as the ball hitting a racket and the ball bouncing on the table (Fig.~\ref{fig:model}B3).
For example, for the events of ball hitting, we computed the ball velocity and the distance between the ball and the players' hands.
\modified{To correctly obtain the poses of the players, we adopted Faster R-CNN for player detection, and filtered and clustered the bounding boxes using k-means for player tracking from both sides.
In the computation of the distance between the ball and the players, sometimes the hand nodes were missed by Openpose because of the occlusion.
To resolve this problem, we additionally considered the neck nodes, which have never been missed by the model during testing.}
We regarded the ball hitting time as the time when the ball velocity changes the direction, and the distance reaches a bottom. 
These potential moments are regarded as anchors, which can help to precisely locate events occurring in a long video.

\subsection{Acquiring Context-level Information.}
\label{sec:interface}
For the context-level information, we designed a user interface to support the calibration of the temporal and spatial attributes of anchors, and the creation of contextual information on the events according to different analytical goals.

The user interface has three major components: the anchors (Fig.~\ref{fig:interface}B, \ref{fig:interface}D), a calibration box (Fig.~\ref{fig:interface}A), and an annotation box (Fig.~\ref{fig:interface}C). Anchors visually present when and where an event occurs.
The calibration box and annotation box support interactive control of anchors, and creation of annotation, respectively.

\textbf{Anchors}
An anchor contains the temporal and spatial information of an event. 
We visualized the spatial attribute ($x, y$) directly on the video frame and the temporal information ($t$) on a timeline. 
Fig.~\ref{fig:interface} illustrates an anchor on an event where the ball hit the table. 
The red point on the table (Fig.~\ref{fig:interface}B) shows where the event happened, for example, where the ball bounced on the table. For the temporal information, we used a highlighted mark on a timeline to show where the event is on the video clip (Fig.~\ref{fig:interface}D).
Different colors of marks on the timeline indicate different mark types. Blacks marks are those that have not been calibrated, and green ones are those that have been calibrated. The red mark is the one that is currently being examined. 

\textbf{Calibration Box}
Anchors are automatically detected by algorithms and inevitably contain errors. The calibration box is used to calibrate the time of an anchor. The design of the calibration box is inspired by "Hotbox"~\cite{kurtenbach1999hotbox}, a menu widget that arranges menu items in a circular manner. We divided the circular box into four functional areas: the left and right areas for correcting the timestamp of an anchor, and the top and bottom areas for adding or removing an anchor (Fig.~\ref{fig:interface}A). When the user clicks the mouse button on a video, the calibration box appears and centered at the cursor. To correct a timestamp, the user can hold the mouse button and drag the cursor left or right to move timestamp backward or forward. If an anchor is useless, the user can delete it by dragging down to the delete area (with a minus symbol). To add a moment as an anchor, the user can invoke the calibration box and drag up to the addition area (with a plus symbol). 

\textbf{Annotation Box}
With the annotation box, the user can interactively create and modify the  annotation of an event. Similarly, the annotation box is also a customized "Hotbox". The number of functional areas is determined by the number of annotation data types. 
Fig.~\ref{fig:interface}C illustrates a scenario where the annotation box is used to annotate the tactics in a rally.

\section{Evaluation}
We conducted two experiments to evaluate how \system{} for Table Tennis (ETT) can assist the annotation of table tennis match videos. The first experiment focused on a task concerning event-level information, and the second on a semantic task at the contextual level. 

\begin{figure*}[tb]
    \centering
    \includegraphics[width=0.8\textwidth]{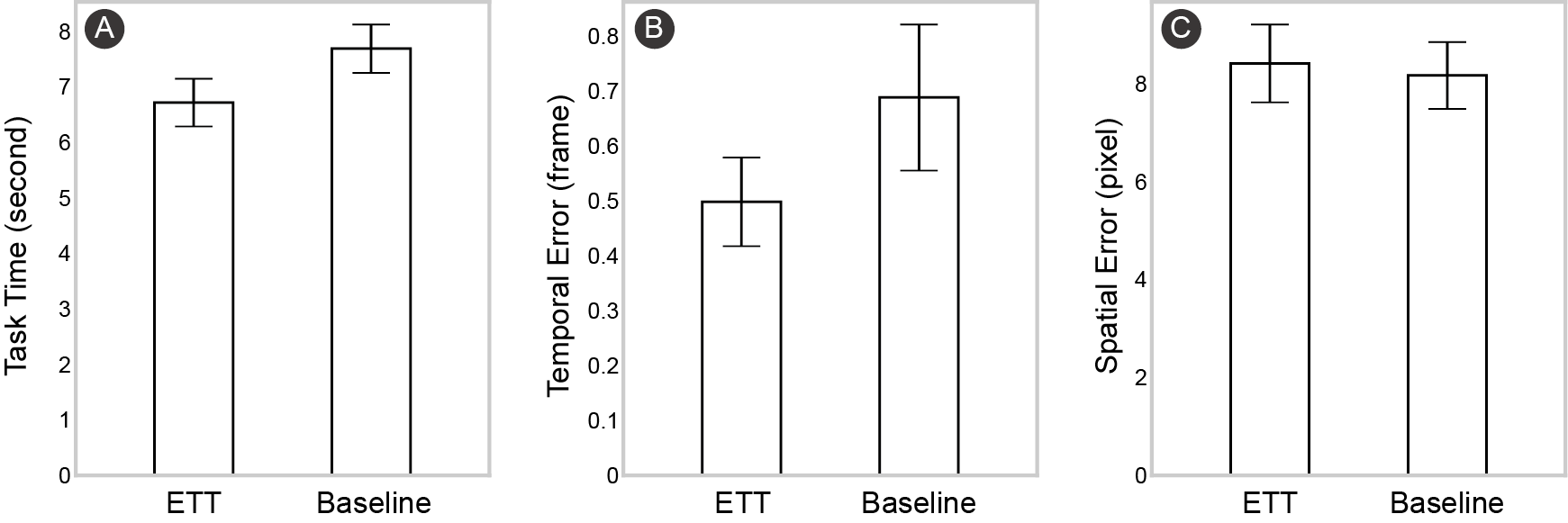}
    \caption{Results of Experiment 1. The bar depicts the mean value and the error bars represent the 95\% confidence interval.}
    \label{fig:experiment_1}
    \Description[Results of Experiment 1]{The results of Experiment 1 are illustrated, where the bar depicts the mean value and the error bars represent the 95\% confidence interval.}
\end{figure*}

\subsection{Experiment 1: Annotating Event-Level Information}
The experiment is a within-subjects design. Two treatments are ETT and a baseline system. 

\subsubsection{Participants}
We recruited 8 (male: 6, female: 2) participants. They all played table tennis regularly (at least twice a week), and knew the sport well. We paid each \$10 for their participation.

\subsubsection{Task}
In this experiment, we asked the participants to find one of the most frequent events: the ball hitting the table in a rally. They were asked to record when and where the ball hit the table in a given video. We chose the final of the \textit{ITTF World Tour 2019} between Ito Mima and Chen Meng. There are 10 rallies in the video, and the length of each rally is between 94 to 175 seconds. The number of the target events in each rally ranges from 5 to 9. 

\subsubsection{Apparatus}
ETT and a baseline system were used. ETT presented the anchors on the timeline, and visualized the spatial position as a highlighted mark overlaying video image. When a video was played, the video slowed down when approaching an anchor, and paused at it. Participants could use the calibration box to adjust its timestamp.
\modified{The baseline system had a structure and an appearance similar to ETT, but with some functions of ETT disabled, including the automatic slowing down and pausing at event timestamp, the calibration box, and the anchor visualization on the timelines.}
To ensure its usability, the baseline system had a video playback tool for participants to watch and control video with a keyboard. When seeing the ball hitting the table, participants could click the position where the ball hit. The system can record mouse-clicking time and location on the video.

\subsubsection{Procedure}
Participants were required to annotate all ten rallies with ETT and the baseline system. Half of the participants used ETT first, and then the baseline. The other half reversed the order.
In each condition, participants went through three steps: training, test, and post interview.
In the training step, they were introduced to the task and the system, and practiced annotation on five rallies different from those used in the test. They could ask any questions about the task and the user interface.

After being familiar with the task and the system, participants took the test. Participants were requested to finish each annotation as fast as they could while ensuring annotation accuracy on time and location. Annotation data was recorded automatically by the system in both conditions. 

After finishing all tasks, they were interviewed for their feedback on tasks and systems. The whole experiment lasted for 30 minutes, 15 minutes for each condition.

\subsubsection{Results}
In total, we collected 461 valid annotations in ETT, and 466 in the  baseline condition. We compared the task time and errors between two conditions (Fig. \ref{fig:experiment_1}). 

\textbf{Task Time} Task time in this experiment was computed as the time difference between annotating two consecutive events. 
The mean times in two treatments are 6.72 seconds ($SD=4.40$) for ETT and 7.69 seconds ($SD=4.49$) for the baseline (Fig.~\ref{fig:experiment_1}A), respectively. The result of a t-test shows that ETT is significantly more efficient than the baseline system in completing the task  ($t=2.49, p=.013$). 

\textbf{Task Errors} We analyzed two types of errors in annotation: temporal and spatial errors. Temporal error was measured as the difference between the frame where a participant annotated and the correct frame, and spatial error was the pixel difference between where a participant clicked and where the ball really hit. The average temporal errors in two treatments are 0.50 frame ($SD=0.86)$ for ETT and 0.85 frame ($SD=1.41$) for the baseline (Fig~\ref{fig:experiment_1}B). A t-test shows the difference is significant ($t=2.52, p=.01$). For the spatial error, the averages are 8.42 pixels ($SD=7.9$) for ETT and 8.18 pixels ($SD=7.04$) for the baseline (Fig~\ref{fig:experiment_1}C), and no significant difference was found between them ($t=0.13, p=.90$). These results indicate that ETT outperforms the baseline in temporal accuracy, and is comparable in spatial accuracy.

\textbf{User Feedback}
In the post interview, all participants, except one, preferred ETT. They mentioned that the anchors in ETT were very helpful, and assisted them to locate the target events more easily, as one participant said: ``\textit{automatically pausing around the events prevents me from missing the event, because the match is at a fast pace}.'' 

\begin{figure*}[tb]
    \centering
    \includegraphics[width=0.8\textwidth]{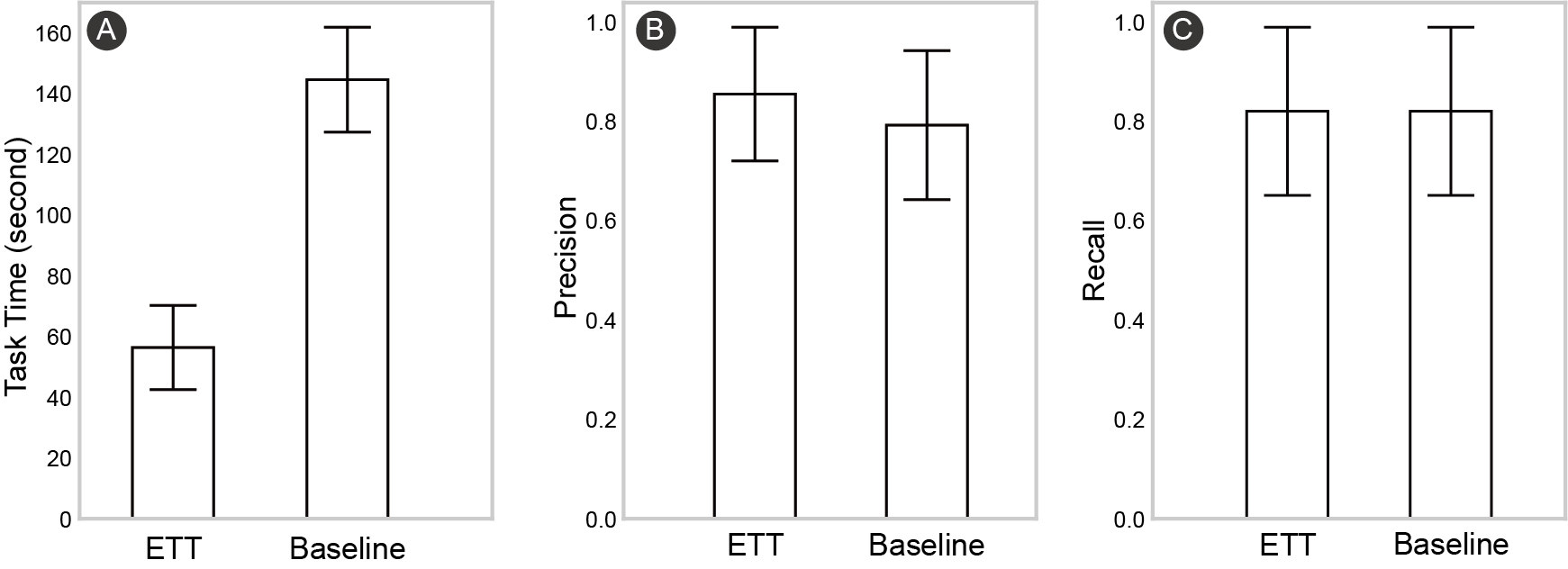}
    \caption{Results of Experiment 2. The bar depicts the mean value and the error bars represent the 95\% confidence interval.}
    \label{fig:experiment_2}
    \Description[Results of Experiment 2]{The results of Experiment 2 are illustrated, where the bar depicts the mean value and the error bars represent the 95\% confidence interval.}
\end{figure*}

Although the user interface of ETT is slightly more complicated than that of the baseline and includes the hotbox design that is less common, most participants were positive about the user interface in general. Two participants indicated that showing the locations of anchors on a timeline helped to improve the efficiency in annotation. One participant commented: ``\textit{this allows me to know in advance how many ball positions need to be labelled, and roughly when to be marked}.'' Some participants were enthusiastic about the hotbox design actually, as three participants indicated that this design was efficient for controlling the video time in annotation.

One participant expressed a concern with errors of annotation in ETT. With the baseline system, participants had to check the video frame by frame to find a specific event. With the suggestions in ETT, however, participants could accept the suggestions from the algorithms without checking whether there was any error in the suggestion. This concern is legitimate, considering the possible errors of computer vision algorithms.

\subsection{Experiment 2: Annotating Context-Level Information}
The second experiment is also a within-subjects design, with two treatments of ETT and a baseline system. The overall design of this experiment is the same as the first experiment. 

\subsubsection{Participants}
We recruited 8 table tennis analysts (male: 5, female: 3) for this experiment. They were all former professional players, and had extensive knowledge of the sport. We paid each participant \$15 for their participation.

\subsubsection{Task}
We asked participants to identify high-level tactics occurring in the final of the \textit{ITTF World Tour 2019} between Ito Mima and Chen Meng. The task was to find the rallies where Ito used the tactic of ``serve and attack" and won the rally. This tactic refers to an approach that the server launches an attack at the third stroke immediately after the opponent receives the ball. 
We chose two games (G1, G2) from the match. Both games contained 24 rallies: one lasted 11 minutes with 2 qualified rallies, and the other 8 minutes with 2 qualified rallies.

\subsubsection{Apparatus}
We used ETT and a baseline system used by professional table tennis analysts. ETT generated a series of anchors of potential rallies, and participants needed to locate and verify these anchors.
They needed to explore all rallies and annotate ``serve and attack'' with an annotation hotbox.
The rule to generate the anchors is that a qualified rally was served and won by Ito and the total strokes by two players were more than 2.
\modified{The baseline system had a structure and an appearance similar to ETT, but with the annotation box and the anchor visualization on the timeline disabled.
Alternatively, there is a confirm button in the baseline system to specify the qualified rally.
Similar to the baseline system in Experiment 1, the features of basic video control as other video players were preserved here.
To annotate a video, participants needed to manually check all rallies one by one, and to click the confirm button for qualified rallies.}

\subsubsection{Procedure}
Participants were required to identify qualified rallies from two games, G1 with ETT and G2 with the baseline system. We could not use the same game in two treatments because participants, as professional analysts, could remember the results from a previous treatment easily. Two games were chosen carefully to make sure the task difficulties on them were comparable. We could not find two games with exactly the same time length, so between G1 (11 minutes) and G2 (8 minutes) we chose G1 for ETT and G2 for the baseline to give the baseline an edge. Half of the participants annotated G1 with ETT first, and then G2 with the baseline. The other half reversed the order.

All participants went through the training, testing, and post interview steps. Videos used in training differed from those in test. The experiment lasted about 20 minutes, 10 for each treatment.

\subsubsection{Results} In total, we collected 16 annotated results: eight in ETT, and eight in the baseline. We analyzed task time and error in two conditions (Fig. \ref{fig:experiment_2}).

\textbf{Task Time} The task time on annotating rally tactics was computed as the time between the start and the end of verifying all rallies in a game. The average times for completing the tasks in two conditions are 56.4 seconds ($SD=21.4$) per game for ETT and 144.5 seconds ($SD=26.7$) per game for the baseline (Fig.~\ref{fig:experiment_2}A). The result of a t-test shows a significant difference between the means ($t=7.30, p<.001$), implying a better efficiency of ETT in support of this complex annotation task, despite the fact that the video annotated in ETT is longer than that in the baseline.

\textbf{Task Errors} We examined the precision and recall of the annotated results.
Precision was computed as the ratio of correct annotations to the total submitted annotations, and recall was the ratio of the correct annotations to the ground-truth.
The ground-truth was produced by one of the domain experts we interviewed (\textbf{E1}) (Section~\ref{sec:interview}).
The average precision is 0.854 ($SD=0.194$) for ETT, slightly higher than that for the baseline, 0.792 ($SD=0.217$) (Fig.~\ref{fig:experiment_2}B). The average recall is the same for both treatments, 0.813 ($SD=0.242$) (Fig.~\ref{fig:experiment_2}C). 

\textbf{User Feedback}
In the post interview, all participants preferred ETT.
They liked the way that anchors help them efficiently locate the potential rallies.
They also enjoyed the user experience in interacting with anchors, as one participant commented: \textit{``the anchors have indicated when Ito will serve and win the rallies, so that I do not have to remember this condition, and just need to focus on the tactic analysis.''}
Another participant added: \textit{``with the help of anchors, I can confirm the tactic type of a rally by only watching the first three strokes.''}

\section{discussion}
The results of our evaluation show that the interaction system based on the proposed \system{} framework can improve the work on annotating table tennis videos. For ordinary users, who may be interested in important movements in a match, this method can help them more quickly identify those events and achieve slightly better annotation accuracy. For experienced analysts, who care more about complex techniques used in a match, the system can improve the efficiency in their work significantly, with similar task accuracy. These results indicate the reliability of our framework in support of such annotation activities, the robust of our computer vision algorithms, and the good usability of our system. 

\modified{
By observing the use of \system{} for annotation, we found that our method can help users overcome some barriers they faced under their old practices. First, from the perspective of interactions, \system{} allows users to focus more on important analytical tasks by freeing them from repetitive interaction tasks. With their old tools, experts have to interact with the keyboard frequently to locate the timestamps of the events before they analyze and annotate them. With our tool, experts have learned that they can trust the pre-computed and filtered timestamps of these events, and can directly focus on judging and recognizing the events. Second, \system{} provides better support for the integration of necessary data with analytic goals. With their old tools, the expert usually divides their whole workflow into two stages, the annotation stage and the analysis stage. The focus of the first stage is on filing video clips and recording such detailed data as stroke type and stroke position. After the completion of such data preparation work, they then shift to the second stage and use different types of data for various analytical tasks. After using \system{}, experts discovered a new annotation-on-demand approach. For example, in Experiment 2, the candidates of the required rally can be filtered and retrieved quickly with the basic information provided by computer vision models. The experts can annotate the detailed attributes of the strokes when necessary. Third, our method can help to reduce the cognition load and shorten interaction processes. Under their old practices, experts have to annotate the stroke attributes at the rally level, because the whole match is clipped into rallies manually in advance. To be efficient, they often try to memorize the attributes of several consecutive strokes and record the results at the same time when watching the rally video. After annotation, they also need to replay and review the whole rally for validation. Sometimes, missing a stroke can lead to several extra replays to discover and correct the errors. Our tool uses computer vision models to provide them with the fine-grained information, so that they quickly and accurately see and obtain required data attributes, such as the timestamps of the strokes. Consequently, they can reserve their valuable cognitive resources for analytical tasks, rather than the memorization of supporting data, and potentially avoid the mistakes caused memory errors and resulting task repetitions.}

Our \system{} can support various statistical and decision-making tasks. Here we provide two scenarios based on the tasks seen in our evaluation study. One scenario concerns the use of our \system{} for accurate statistical analysis by leveraging crowdsourcing. In table tennis, analyzing ball positions in a full match statistically requires a dedicated analyst to mark the exact positions of the ball on the table. It usually takes 30 to 40 minutes to complete the task. With the help of our system, this task can be accomplished by distributing individual video segments of ball landing, which are generated by computer vision algorithms, to crowd workers. Verifying and calibrating a ball position is an ideal crowdsourcing task, because of the short time required to do it, about 6.72 seconds, as we learned from our study (Fig.~\ref{fig:experiment_1}A), and no requirement for domain knowledge. 
Another scenario is related to quick decision-making that involves domain experts. In real matches, coaches and players often need to adjust their tactics or strategies based on the performance of the opponent. Our system can help them quickly search through videos to find the tactics of the opponent and make the necessary adjustments on tactics or strategies. As our experiment results show that the average time to discover the rallies with a specific condition is less than a minute (56.4 seconds) for domain experts (Fig.~\ref{fig:experiment_2}A), our system can provide real-time support for coaches and players during a break between rallies. Thus, our system can potentially change the ways people conduct video analysis by reducing the requirement on domain knowledge, or by allowing the use of video data for decision-making in fast-paced situations.

Although the effectiveness of our framework is demonstrated through a system for table tennis annotation, this approach can be applied to video annotation in other racket sports. For racket sports like tennis and badminton, with similar image setups and structures in broadcasting videos, our framework can be directly applied, with proper algorithm training. For other racket sports, such as racquetball and squash, more work is needed to refine computer vision algorithms to adapt to the different image structures and player movement patterns in videos, but our framework to anchor analysis to events can still be used, because many rules of these sports, such as those concerning ball hitting a racket and the court, are similar to those of tennis, table tennis, and badminton. 

There are some limitations in our work. First, our work could be more flexible on the definition of events. Our current definition of events as ball-object contact works well for table tennis analysis, but people may be interested in other events, such as sudden movement changes of a player. One approach to expand the definition of events is to develop an event syntax that includes essential elements (e.g., ball, player, court, net, racket, etc.), their attributes, and the spatial and temporal relationships among them, and then let users interactively define a new event type under the syntax. 

The second limitation is insufficient use of audio from videos.  Audio information could be used for event recognition by detecting the sound of ball contact(e.g., stroke detection~\cite{ono2019historytracker}), and by adding information from another sensory channel, provide users with additional information for annotation and make data annotation more engaging. 

Furthermore, we need better mechanisms to motivate users to carefully examine and calibrate the results suggested by algorithms. As shown in Fig.~\ref{fig:experiment_1}C, the spatial error in annotating ball position with ETT is slightly larger than that with the baseline system, although the difference is not found significant. New designs are needed to encourage user engagement with algorithms and computational results. 

\section{conclusion}
This paper proposed \system, a framework to support data annotation of racket sports videos. This framework uses events recognized by computer vision algorithms as anchors to help users locate, analyze, and annotate objects more efficiently. Based on this framework, we implemented a system for table tennis annotation, and the results from our evaluation study on the system show significant improvement of user performances in simple annotation tasks (e.g., labeling ball position) and in complex tasks requiring domain knowledge (e.g., labeling rallies with specific tactics).   

Our method can guide the design of systems for video annotation in other racket sports, such as tennis and badminton. With improvements on algorithms and interaction designs, its application domain can be extended. We will explore designs that allow users to define new event types, so that the system can recognize and process more complex events and support annotation on fast and dynamic videos in other domains. In addition, we will improve interaction design to make users more engaged with computational results, further strengthening the collaboration between human brain powers and machine computation powers. 

\section{Acknowledgments}
\modified{We thank all participants and reviewers for their thoughtful
feedback and comments. The work was supported by National Key R\&D Program of China (2018YFB1004300), NSFC (62072400), Zhejiang Provincial Natural Science Foundation (LR18F020001), and the 100 Talents Program of Zhejiang University. This project was also funded by the Chinese Table Tennis Association.}

\bibliographystyle{ACM-Reference-Format}
\bibliography{main}

\end{document}